\documentclass[manuscript]{emulateapj}

\slugcomment{Submitted to The Astrophysical Journal}

\shorttitle{Asteroseismology of HD 17156}
\shortauthors{Gilliland, et al.}
\def\rhomean{\langle \rho_* \rangle}
\def\CF{{\cal F}}
\def\note #1]{{\bf #1]}}
\def\Msun{\,{\rm M}_\odot}
\def\CM{{\cal M}}

\begin{document}


\title{ASTEROSEISMOLOGY OF THE TRANSITING EXOPLANET HOST HD 17156 WITH HST FGS\altaffilmark{1}}


\author{Ronald L. Gilliland\altaffilmark{2}, Peter R. McCullough\altaffilmark{2} and
Edmund P. Nelan\altaffilmark{2}}
\affil{Space Telescope Science Institute, 3700 San Martin Drive,
Baltimore, MD 21218}
\email{gillil@stsci.edu}
\author{Timothy M. Brown\altaffilmark{3}}
\affil{Las Cumbres Observatory Global Telescope, Goleta, CA 93117}
\author{David Charbonneau\altaffilmark{4} and Philip Nutzman\altaffilmark{4}}
\affil{Harvard-Smithsonian Center for Astrophysics, 60 Garden St., Cambridge, MA 02138}
\author{J{\o}rgen Christensen-Dalsgaard\altaffilmark{5} and Hans Kjeldsen\altaffilmark{5}}
\affil{Department of Physics and Astronomy, Aarhus University, DK-8000 Aarhus C, Denmark}


\altaffiltext{1}{Based on observations with the NASA/ESA {\em Hubble
Space Telescope}, obtained at the Space Telescope Science Institute,
which is operated by AURA, Inc., under NASA contract NAS 5-26555.}


\begin{abstract}
Observations conducted with the Fine Guidance Sensor
on {\em Hubble Space Telescope (HST)} providing
high cadence and precision time-series photometry were obtained over 10 consecutive days
in December 2008 on the host star of the transiting exoplanet HD 17156b.
During this time 1.0$\times 10^{12}$ photons (corrected for detector
deadtime) were collected in which a noise level of 163 parts per million
per 30 second sum resulted, thus providing excellent sensitivity to 
detection of the analog of the solar 5-minute p-mode oscillations.
For HD 17156 robust detection of p-modes supports determination of 
the stellar mean density of
$\rhomean = 0.5301 \pm 0.0044$\,g\,cm$^{-3}$ from a detailed
fit to the observed frequencies of modes of degree $l = 0$, 1, and 2.
This is the first star for which
direct determination of $\rhomean$ has been possible using both asteroseismology
and detailed analysis of a transiting planet light curve.  Using the density
constraint from asteroseismology, and stellar evolution modeling
results in
$M_* = 1.285 \pm 0.026 \Msun$, $R_* = 1.507 \pm 0.012 {\rm R}_{\odot}$,
and a stellar age of $3.2 \pm 0.3$\,Gyr.
\end{abstract}





\objectname{HD 17156}

\section{INTRODUCTION}

General knowledge of stellar structure and evolution for normal
main-sequence stars has long been a mature topic, with forefront
research driven by increasingly precise observational constraints,
and theoretical investigations including effects such as heavy 
element diffusion and overshoot of material at convection zone boundaries.
Transiting planets, the first of which HD 209458 was discovered only 
a decade ago (\citealp{cha00}; \citealp{hen00}) have provided a wealth
of detailed physical information about dozens of planets.  Of more 
relevance to the current paper, high quality transit light curves also
return a direct determination of the mean density of the host 
star, $\rhomean$, independent of stellar evolution models (e.g.
\citealp{sea03}; \citealp{soz07}, and \citealp{win08}). 
Asteroseismology has long promised fundamentally new and precise
constraints on stellar structure and evolution (e.g. see review of
\citealp{bro94}), and with the use of state-of-the-art ground-based
spectroscopic radial velocities prompted largely by the burgeoning
efforts for RV planet discovery has delivered several successes
in recent years (e.g. see review of \citealp{bed08}).  
The advent of space-based photometric missions promises to provide
robust results from oscillations on a much larger number of stars,
as hinted at by the early successes from CoRoT (\citealp{mic08}).
Detection of several low-angular-degree p-modes also provides a 
direct constraint on $\rhomean$ as will be discussed at length
for HD 17156 herein.

In this paper we present photometry of the host for the unusually interesting
exoplanet HD 17156b given its long orbital period of 21.2 days
discovered by \citet{fis07} in a Doppler survey, and shown to 
have transits by \citet{bar07}.  The nearly 10 days of high precision
{\em HST} photometry suffices to obtain secure detection of  
some 8 individual p-modes, an accurate determination of the asteroseismic large
separation, and hence accurate measure of the stellar mean density.
A companion paper by \citet{nut09} will present analyses of 
complementary {\em HST} observations through three separate transits
of HD 17156b to fix $\rhomean$ from transit light curve analysis.
This provides the first instance of obtaining such precise measures
of this fundamental stellar parameter from two entirely different 
techniques, thus providing not only the intrinsically interesting
physical measurement and associated interpretations, but also 
providing a test of the two methods.  We find a gratifying level
of consistency from independent application of the two techniques,
thus providing enhanced confidence that both perform as expected.

Selection of HD 17156 as the target and the observations obtained
for the asteroseismology part of this project are discussed in \S 2.
Extensive, and unique, procedures invoked for the analysis of 
{\em HST} Fine Guidance Sensor time-series photometry are discussed in \S 3.
The evidence for, and detailed results of power spectral and related
analyses for stellar oscillations are covered in \S 4.
\S 5 presents interpretation using stellar evolution and eigenfrequency
analyses.
Comparisons with the transit light curve analyses are provided in \S 6
along with a look to the future prospects for similar results on 
a much larger number of systems from the {\em Kepler Mission}.

\section{TARGET CHOICE AND OBSERVATIONS}

\subsection{Target Choice and Background on FGS Use for Photometry}

The underlying motivation of this study, anticipated for years in 
application with the {\em Kepler Mission}, was to simultaneously 
challenge the two relatively new techniques of transit light curve
analysis and asteroseismology by comparing results for
$\rhomean$ on the same object, while at the same time providing a 
new benchmark of accuracy for stellar and planetary quantities.
Application of asteroseismology, with well-posed observations, 
promises determinations of stellar density to $<$1\%, and ages
to $<$10\% -- very desirable constraints on stellar structure 
and evolution.  However, to reach such results requires a
large allocation of observing time, both to reach the necessary
precisions of better than a meter per second if through radial
velocities, or near 1 part-per-million (ppm) if through broadband
photometry, and to reach the necessary frequency resolution of
order 1\,$\mu$Hz.  For photometry these constraints essentially
come down to requiring that of order $10^{12}$ photons be collected
over a period spanning about 10 days, and for which near Poisson
limited results can be maintained.  The best ground-based
attempt with photometry to date involved using a longitude-distributed network
of 4-m telescopes for a one week period in 1992 \citep{gil93}
to find oscillations in a cluster of slightly evolved stars
expected to have favorable amplitudes of a few times solar; although
a technical success this study failed to detect oscillations.
The capability of {\em HST} to provide successful asteroseismology
of solar-like stars has long been expected to hold, but the 
necessity of dedicating 10 days of observations to one bright 
star has not supported successful applications.  In the fall of 
2008 {\em HST} had lost the use of its primary instruments,
ACS, NICMOS and STIS, leaving only the
imager WFPC2 with nearly 15 year on-orbit in service.  In addition, Side A of the 
{\em HST} electronics communication package required to utilize WFPC2 failed, thus
precipitating a delay in the much anticipated recent servicing
mission to upgrade Hubble's instruments.  When use of Side B 
did not initially succeed, we submitted this program to allow
good use of the remaining science capability of {\em HST} at that time: the
Fine Guidance Sensors \citep{nel08}.  The FGS can provide exquisite interferometric
position determinations, and time series photometry through 
summing the counts of the 4 photomultipliers (PMTs) on each FGS
in POSition mode, as well as high angular resolution, narrow field
interferometry with TRANSfer mode observations, see \citet{nel08} for 
in-depth discussion of general FGS capabilities.

Our application for DD time on {\em HST} was granted as GO/DD-11945,
``Asteroseismology of Extrasolar Planet Host Stars".
Our photometric requirements from the FGS exceeded those reached 
in its previous use, which usually involved single-orbit visits
timed to coincide with ingress or egress of planet transits
(\citealp{wit05}; \citealp{bea08}).  Extensive experience has
shown with other instruments on {\em HST} that during the 
first orbit, as the spacecraft thermally adjusts at a new 
orientation, photometric stability is much lower than in successive orbits.
We were initially granted a test block of 8 contiguous orbits 
to coincide with a transit of HD 17156b.  As we expected, after the
first orbit much better stability followed and these observations
are discussed and used in \citet{nut09}.  With this successful
demonstration of FGS capability we were granted the 150 orbits
necessary to follow HD 17156 for 10 days during which it was
in the Continuous Viewing Zone (CVZ) of {\em HST} for much of this time.

The choice of HD 17156 as a target followed from a review of
all 42 then known transiting planet host stars to see which
would hold the best prospect of asteroseismology with a 10 day
block of dedicated {\em HST} time with FGS.
Adopting stellar parameters from Frederic Pont's table at
http://www.inscience.ch/transits (no longer active -- the equivalent,
albeit updated information may be found at http://exoplanet.eu)
and scalings for expected
oscillation properties \citep{kje95} resulted in a prediction
of amplitudes at 6.9 ppm for HD 17156.
With 10 days in the {\em HST}
CVZ a signal-to-noise of $\sim$ 7 was predicted for the highest
amplitude modes (and ultimately close to the value reached)
for these FGS observations.
Since this S/N for p-mode detections was well above that predicted for any other transiting planet
host, and the system is unusually interesting given its 21.2 day 
period compared to the more common 3 -- 5 days of Hot Jupiters
studied to date, the target choice was clear.

Although the discovery paper for HD 17156 appeared only three years
ago, the system is now very well studied.  In addition to its 
21.2 day orbit, the planet has an eccentricity of 0.68.
HD 17156 is one of just three transiting extrasolar planets for 
which Rossiter-McLaughlin radial velocity observations through
transit have determined the relative stellar rotation and orbital
plane alignments to good accuracy (\citealp{coc08}; \citealp{nar09}).  For HD 17156
the stellar rotation and orbit of HD 17156b are well aligned. 
Asteroseismology holds the promise of also providing information
on the stellar rotation angle relative to the plane of the sky
when rotational splitting of individual nonradial oscillation
modes can be detected.  
For HD 17156 the Rossiter-McLaughlin study of \citet{nar09}
fixes $v \sin i$ at $4.07 \pm 0.28$ \,km\,s$^{-1}$, which coupled
with the estimate of $1.45 \, {\rm R}_{\odot}$ for HD 17156 (\citealp{win09})
leads to an estimate of the stellar rotation period of $18.6 \pm 1.4$
days.  Since our observations span only half the expected rotation
period, rotational splitting of nonradial mode frequencies will
remain under our frequency resolution.  
(In principle the full splitting for $l$ = 2, $m$ = $\pm$2 modes
could be over-sampled by a factor of two with our window, but our
marginal signal-to-noise seems unlikely to support such higher
order interpretations.)
HD 17156 has been well studied spectroscopically (\citealp{fis07};
\citealp{amm09}), via transit light curve studies that determine
stellar density, and augmented with stellar evolution comparisons
the stellar mass and radius separately (e.g. see most recent 
such study by \citealp{win09}), and has a parallax from {\em Hipparcos}
with a relative error of only 5\% (\citealp{lee07}).
Table 3 will include a summary of $T_{\rm eff}$, $\log(g)$, and [Fe/H]
from the spectroscopic studies, as well as the primary results
from this paper.
Not only is HD 17156 the best star for these {\em HST} 
observations in a technical sense (CVZ, high count rate possible,
and `large' predicted oscillation amplitudes),
it is also near the absolute top of transiting extrasolar
planet hosts in terms of intrinsic scientific interest.

\subsection{The Observations}

Observations of HD 17156 with the Fine Guidance Sensor 2 on {\em HST}
were obtained for 147 contiguous {\em HST} orbits spanning 9.67 days
over 22 -- 31 December 2008.
Use of FGS2r, which had not previously been used for science observations,
and was scheduled for replacement (and now has been replaced) in
servicing mission \#4 was selected out of extreme caution at subjecting
the photomultipliers to a summed count level of order $10^{12}$ photons
as would follow for this V = 8.172 star, since some level of lost 
sensitivity is expected proportional to total source exposure.  
In the end these observations were measured to have reduced the FGS2r
sensitivity by a quite acceptable level of 0.2 -- 0.3\%, suggesting
that the better calibrated FGSs could have been safely used.
HD 17156 is only 0.6 magnitudes fainter than the level at which the
FGS2r digital count registers would saturate, rendering any observations
of a brighter target at full aperture impossible (\citealp{gil09f}).
Since FGS2r had not previously been used for science observations,
and since none of the FGSs had been carefully calibrated for detector 
deadtime a few additional orbits were allocated to this project for
the purpose of calibrating the FGS2r deadtime, background, sensitivity
and noise properties (see results in \citealp{gil09f}).

The Fine Guidance Sensors use photomultipliers (4 in each FGS) located
behind first a beam splitter based on the two senses of linear polarization,
and then behind two Koesters prisms that generate an interferometric signal.
The primary use of the FGSs are to track the pointing of the telescope
on guide stars and provide feedback into the pointing control system.
Used as a photometer in POS mode the observing FGS (FGS2r in this case)
acquires and tracks the target in Finelock, while the two guide
FGSs keep the target stably positioned to start with.

The FGS data is collected at 40 Hz in each of the four PMTs per FGS.
Examined at 40\,Hz the relative counts from the PMTs behind each Koesters
prism show strong (of order one percent) variations with characteristic
timescales of about a second.
The sum of the two PMTs for each polarization remains much more constant
on these short timescales.
We used the F583W filter covering 435 -- 715 nm at full aperture
which provides the maximum throughput available from the FGSs.

After applying a deadtime correction averaging 11.0\% the mean number of counts
per 30 seconds summed over all 4 PMTs on HD 17156 is 4.96 $\times \, 10^7$.
Over the 9.67 days of observing a total of 20208 time series points
were extracted yielding a total count level of 1.00 $\times \, 10^{12}$
photons for which an expected sensitivity limit of 1 ppm would follow 
if the observations are limited only by Poisson statistics.

The observations after summing over the 4 separate PMTs in successive
blocks of 1200 samples at 40\,Hz are shown in Fig. 1.  
The data were transformed to relative photometry by dividing each
sum by the global average.
There are several features in the data cadence to call attention to.
The {\em HST} orbital period often results in gaps.  
During the central 6 days of these observations when the target was
in the CVZ for {\em HST} these gaps result from passing through the 
South Atlantic Anomaly (SAA) when high particle-event rates require the 
FGS photomultiplier high voltage (HV) to be shut down.  Toward the start and end
of the 10 days gaps also follow from occultations from the Earth
as the CVZ viewing is lost.  
True CVZ blocks can be seen in the 2nd through 7th days of Fig. 1
with up to 9 consecutive {\em HST} orbits of unbroken data collection.
A gap occurs over HJD = 27.05 -- 27.33 when for unknown reasons
(particle hit in electronics perhaps) lock was lost and FGS2r 
ceased observation of HD 17156.  During this 6 hour period FGS1 and FGS3 
continued to track their guide stars, and FGS2 observed sky, or at 
least a close approximation of such about 26 arcseconds from the star --
this would prove useful in calibrating corrections for sky background.
Time series points for HD 17156 have only been used from periods
when data quality flags show FGS2r remained in fine lock.
The first data point is at HJD 22.63362 (with leading 2454800 suppressed)
with the last at HJD 32.30393 for coverage of 9.67 days during which
data was collected.  The duty cycle (fraction of total time during
which photons were usefully collected) during this interval was 72.6\%
reflecting the high observing efficiencies possible with {\em HST}.
Clearly the data structure will introduce aliases in power spectra
associated with the orbital period of {\em HST} and the one per day
cycling of CVZ periods unbroken by the SAA.

\begin{figure}[h!]
\resizebox{\hsize}{!}{\includegraphics{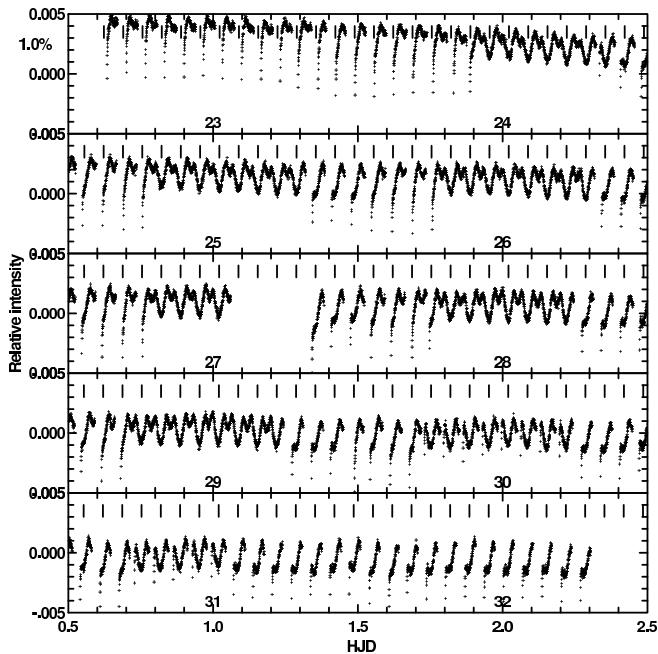}}
\caption{Time series photometry for 30 second sums over the full ten days
with two days per panel shown starting at the top and progressing
to the bottom.  The numbers within the panels show start values
for HJD days 23 -- 32 (-2454800).  Each panel has the same 1.0\% full range,
which to help guide the eye is also printed at the upper left.
See text for further details.
\label{fig1}}
\end{figure}

The goal of this project is to detect several independent modes
of oscillation on HD 17156, the highest amplitude of which is expected
to be $\leq$ 10 ppm, and to obtain a S/N of several per mode in doing so.
The simply extracted time series shown in Fig. 1 show variations 
peak-to-peak at a range of 1\%, or 1000 times larger than the peak intrinsic
variations expected from the star.
Fortunately (and by design, we would not have attempted this otherwise)
the stellar oscillations for HD 17156 are expected to have periods of
6.5 -- 13 minutes, i.e. well separated from the bulk (but not all)
of characteristic time scales reflected in the systematics apparent
in Fig. 1.  The next section will explain the rather extreme reduction
steps taken to bridge the gap between simply obtained, and necessary,
noise levels in these data.

\section{REDUCTION OF FGS2r ASTEROSEISMIC DATA}

\subsection{Removal of {\em HST} Orbital Artifacts}

We have experimented with starting the data reductions using 
the individual 40\,Hz level counts from each of the four contributing PMTs.
Having found no advantages in considering the highest frequencies,
we have settled on analyzing 30 second (1200 samples at 0.025 second) sums.
We have also explored analyzing the pairs of PMTs separately, and 
again no useful utility in doing so was uncovered.

The only exception to working with 30 second sums over all four PMTs
is that we have applied the deadtime correction for the FGS electronics
separately for each PMT at the 40\,Hz level, and subtracted the relatively
unimportant detector background count level of $\sim$3 per sample per PMT
(\citealp{gil09f}).
The equation adopted
for deadtime is (see \citealp{gil09f} and references therein):
\begin{equation}
C_T = C_M/(1.0 - C_M (T_D/T_I))
\end{equation}
where $C_T$ is the true number of counts within the sampling
interval of $T_I$ = 0.025 seconds, $C_M$ is the recorded number
of counts in this interval, and the $T_D$ have values of 
210.6, 306.0, 260.3 and 286.0 nano-seconds for the four PMTs.
In a general sense the correction for detector deadtime is 
relatively unimportant in this paper, but essential for the 
analysis of transit depths since both the absolute count level
and changes to that scale with this correction.

Continuing discussion of reductions by reference to Fig. 1,
note that after each gap in the observations there is a rapid 
rise in the relative intensity.
The sequence within each new observation (single orbits
of {\em HST} normally, up to 9 consecutive orbits when in the 
Continuous Viewing Zone) is that the two guide FGSs turn on
and establish ``fine lock" on guide stars, and only then is 
the High Voltage (HV) turned on for the science FGS and
fine lock guiding established on HD 17156.
This results in 40 Hz time series on HD 17156 starting
at zero before HV is turned on, a rapid ($\sim$0.1 second)
transition from zero to $\sim$97\% of full counts at 
HV turn on and an exponential ramp up to the full count
level over $\sim$5 minutes.  To facilitate correcting the 
data during the $\sim$5 minutes of stabilization at the
start of each observing sequence, of which we have 101
occurrences in our 10 days, the 30 second data sums are
always started the same number of 40\,Hz steps after HV
turn on as evidenced by rapidly rising counts.  The offset
between HV turn on and the start of the first 30 second
sum has been chosen as 21.5 seconds, a somewhat arbitrary
offset that puts the first 30 second sum at about 0.995
of the ultimate count rate.  More on treating the ramp
up of counts is provided later.

A first {\em HST} orbit which included brief TRANS mode observations
to search for companions to HD 17156,
and the start of time series observations in POS mode has
been discarded, and is not shown in Fig. 1.
The TRANs mode observations show that HD 17156 appears to be a 
point source to the resolution of FGS2r, thus a stellar companion
with a projected separation greater than about 0.015 arcsecond
with a $\delta V$ smaller than about 3 can be excluded.
Since HD 17156 has a Hipparcos distance of $\sim$78 pc \citep{lee07},
0.015 arcseconds corresponds to $\sim$1.1 AU.  The published RV data
do not suggest a stellar mass body with period less than a few years,
so between FGS and RV, companions with $\delta V \, \leq $ 3 are excluded.

A number of features are immediately obvious from 
inspection of Figure 1:  (a) The data values trend 
downward by about 0.3\% over the 10 days.  This has been
separately established to be instrumental, at least at
a precision level of $\sim$0.1\% by observing the FGS
standard Upgren 69 before and after the 10 days which
showed a similar drop (\citealp{gil09f}).  A minor degradation of PMT
sensitivity from exposure to $10^{12}$ photons was expected.
(b) The {\em HST} orbital timescale is shown by successive
vertical tick marks ($95.9184 \pm 0.003$ minutes) to 
guide the eye (and later reductions).  Within {\em HST} 
orbits the time series shows similar waveforms of 
typical full amplitude $\sim$0.025\%.  These waveforms apparently
evolve slowly over the 10 days.  (Not shown -- the two 
PMT pair sums show different orbital waveforms, with 
the variations in one being about twice the other, but
both seem to show equally consistent orbit-to-orbit 
waveforms.) (c)  At the start of each contiguous block
of data the time series show a ramp up of full 
amplitude $\sim$0.4\% following the HV turn on.

The initial approach to data reductions,
as had apparently worked quite well for the more limited
(and simpler orbital waveform) data during the Nov. 7, 2008
transit was based on attempting decorrelations against the x,y
pointing records from all three FGSs, and the ratios of
counts between PMT pairs in FGS2r itself.  These are not
shown; it quickly became evident that large features
appear in the potential decorrelation vectors that do not
appear in these times series data over 10 days.

We will next show plots similar to  Fig. 1
that show several decomposition terms for the obvious artifacts.
These will include in Fig. 2 a minor additive correction
made for varying sky background.
Fig. 3 will document a slow variation in time, picking up time scales
that are long compared to the {\em HST} orbit.  Fig. 4 will 
show an HST orbital wave form that is allowed to evolve 
slowly in time.  The HV ramp up in counts will be detailed in
Fig. 5.  The factors for the slow variation in 
time, the orbital waveform, and the HV ramp are solved
for in an iterative, least squares procedure.
Fig. 6 will show a final tweak that is derived after the
above iterative solution.  We next detail these corrections.
These corrections are applied to the raw time series such 
that subtraction of the Fig. 2 term and a point-by-point division by the four terms from
Figures 3, 4, 5 and 6 will take the raw values of Fig. 1
into the corrected values of Fig. 7.

\subsubsection{Subtraction of sky and particle backgrounds}

As noted in \S 2 the data gap over HJD = 27.05 -- 27.33 
proved useful for deriving a proxy for variations in the
sky background.
FGS1 and FGS3 both observed stars much fainter than our 
target HD 17156.
In order to keep the bearings in the FGSs lubricated the
guide FGSs were rotated between stars -- FGS1 alternated
between two at $V$ = 11.50 and 11.53 ($\times$21.5 fainter
than HD 17156) and FGS3 alternated between $V$ = 12.97,
13.48, 14.31 stars ($\times$83 -- 286 fainter than HD 17156).
Changes of background occur for two reasons:  (1) variations
in the background light, e.g. scattered from the bright 
Earth limb, and (2) changes in the charged particle flux
inducing counts in the PMTs.  Complete discussion developing
FGS1 and FGS3 count rate changes as a proxy for FGS2r need
not be recounted in detail here -- FGS2r is no longer on 
the telescope and was only used for science observations
for the current program.  Suffice it to note that minor
differences in scaling for each of FGS1 and FGS3 to FGS2r
background changes exist, and these are slightly different
between contributions from light and charged particle induced
counts.  Also complicating this, one of the FGS1 stars was
an obvious variable, but for which the intrinsic variations
could be filtered out.  Fortunately, the 6 hour gap 
while FGS2r was observing only sky provided good data for
calibrating use of FGS1 and FGS3 as proxies, and the
implied corrections as shown in Fig. 2 are very minor
for the extremely bright HD 17156.
The subtraction for sky is performed
before any of the more important corrections to be detailed next
are developed in an iterative sense.
Only the occasional brush with outer
SAA regions leads to spikes that still remain under 0.09\%,
and these are well determined.  We believe that uncertainties
in the sky subtraction are quite unimportant.

\begin{figure}[h!]
\resizebox{\hsize}{!}{\includegraphics{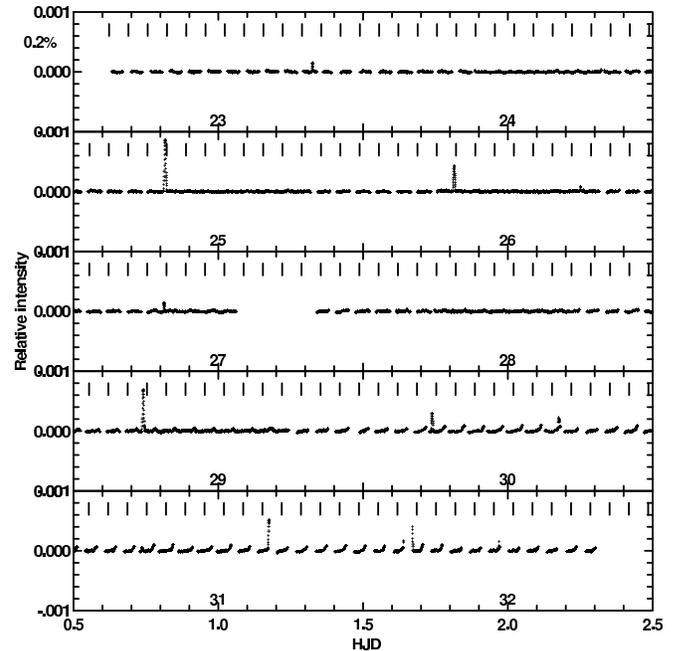}}
\caption{Sky plus counts from charged particles shown as
corrections relative to the HD 17156 count level.  See text
for how FGS1 and FGS3 counts are used as proxies for this.
The sharp, positive spikes result from grazing the South
Atlantic Anomaly.
\label{fig2}}
\end{figure}

\subsubsection{Slow variation of count rate}

The slowly varying count rate is shown in Fig. 3 and
has been derived by running a median filter of full width
equal to 2.2 {\em HST} orbits over the data of Fig. 1 (this
is done only after sky subtraction and having accounted for the orbital wave
form and HV ramp terms to be discussed next).  This filtering should have virtually 
no impact on stellar oscillations for which we expect
time scales of 5-15 minutes in HD 17156, while the filter
is 211 minutes wide.

\begin{figure}[h!]
\resizebox{\hsize}{!}{\includegraphics{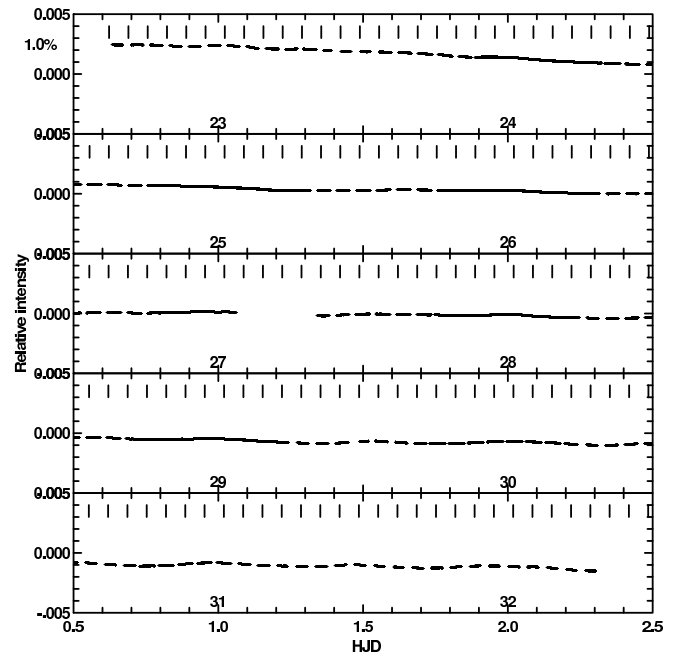}}
\caption{A slowly varying long-term correction factor
derived by running a median filter with width of 0.15 
days over the data.
The 0.2 -- 0.3\% decline overall likely follows from lost FGS
PMT sensitivity, and not intrinsic variation of HD 17156.
\label{fig3}}
\end{figure}

\subsubsection{HST orbit induced orbital waveform}

The most interesting and important artifact in the data
to be dealt with arises from highly repeatable systematics
tied to the {\em HST} orbit.  The source of these variations 
is treated as unknown for these purposes, and it is unknown.
The orbital
waveform derived from the data is shown in Fig. 4.
To derive this an assumed {\em HST} orbital phase is generated
for all the data points (the {\em HST} orbital period is allowed
to change later as needed to produce optimal results, but
a good value is easily derived from inspection).  The 
waveform is solved for on each of 1000 phase points spanning an
{\em HST} orbit with weights set as $\exp (-( \delta t /10.4)^2)$
where $\delta t$ is the time in seconds that the center
of given 30 second sums is out of phase alignment for the
orbital phase point being solved for.  Essentially, one makes a stack of
all the data points folded on the assumed {\em HST} orbital 
period and makes a weighted average at each phase point as
the ratio of the dot product of data and weights divided
by the sum of the weights.  In practice this
solution either does not (initially) use the points impacted
by HV turn on, or (later) includes these points after 
correction for the HV ramp.  Likewise, the solution assumes
the slow variation shown in Fig. 3 has first been removed.
The apparent slow variation in time of the orbital 
waveform is included by forming not a simple weighted sum
at each phase point, but rather a quadratic fit over 
the 10 days in time.  Later in the iteration cycle this is
increased to fitting a cubic in time.  Finally, with the
orbital waveform in hand, which consists of the zero point,
linear, quadratic and cubic polynomial terms at each of 
1000 phase points, interpolation is used to provide the 
correction at the center of each 30 second data point.
We do not claim that this nested, iterative solution  for 
the orbital waveform provides either a unique, or optimal correction.
However, inspection of Figures 1 and 4 shows that the derived
orbital correction does an excellent job of matching 
features, and thus removing the {\em HST}-orbit induced systematics.

a
\begin{figure}[h!]
\resizebox{\hsize}{!}{\includegraphics{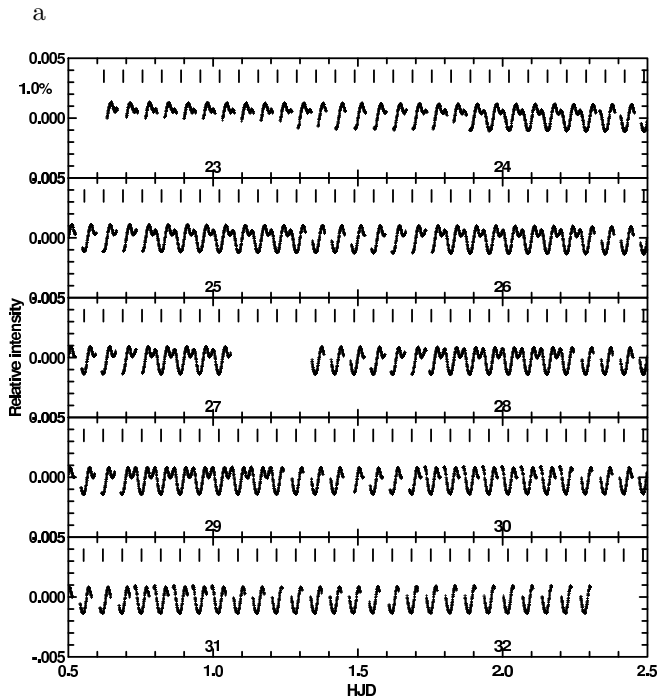}}
\caption{Orbital waveform component of data corrections
plotted on the same times and scale of Figure 1.
\label{fig4}}
\end{figure}

\subsubsection{High Voltage ramp up}

The correction factor for the HV ramp period is shown
in Fig. 5.  As discussed earlier each 30 second sum
is initiated (to $\sim$0.01 second {\em rms}) at exactly the
same offset following HV turn on.  To first order we
assume that after each HV turn on successive data points
experience the same suppression of count rate.
But we can do better.
The offsets in successive accumulated sums after HV turn on show a
strong correlation against the length of time the HV was off,
equivalently the length of time the PMTs and FGS electronics were
not experiencing a high photon flux. 
After removing the linear correlation with time gap a minor
dependence on time over the 10 days is also clear.  In practice the solution
for each successive HV turn on point is formed as a 
multi-linear regression over the time gap, and time
(ignoring only the case following the large 6 hour time gap
which is not consistently off by this amount).  
This solution is followed for the first 135 points in 
sequences, although most of the effect is 
restricted to the first 5-8 minutes.  Following all 
corrections the scatter of points near the start of 
observing sequences is no larger than the general scatter in the time series.
The HV ramp is corrected with essentially no 
residual error, thus restoring some 12 hours of data
that would otherwise need to be dropped
since the variations in these time periods contain frequency
components that would directly affect the ability to cleanly
detect oscillations of 5 -- 15 minutes.

a
\begin{figure}[h!]
\resizebox{\hsize}{!}{\includegraphics{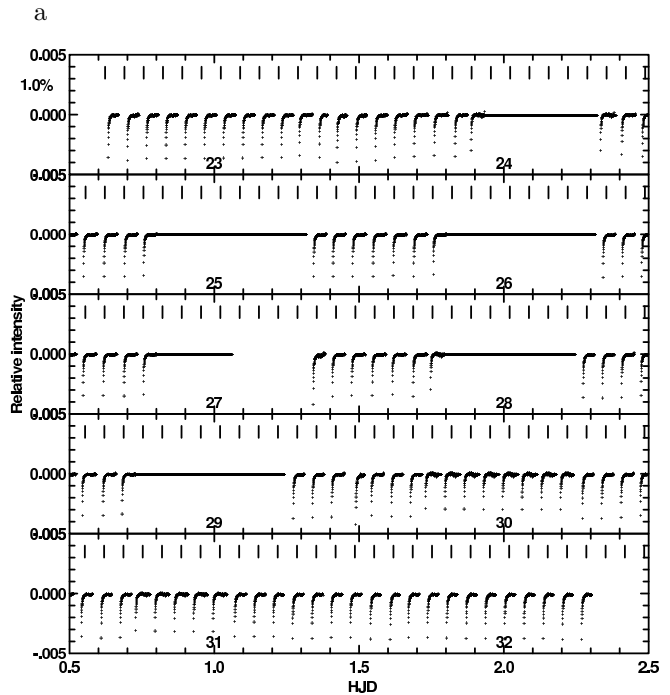}}
\caption{The correction factor for HV ramp up, based
on a bi-linear regression over the preceding time gap, 
and overall time for each successive point in observing 
sequences.
\label{fig5}}
\end{figure}

\subsubsection{Residual orbital corrections}

After having applied the several correction factors discussed
above, it was clear from inspection that some obvious,
albeit much smaller, residuals remained.  
A quadratic fit to each individual data
segment has been effective in dealing with much of the 
residuals.  Before applying this correction the overall
{\em rms} had been reduced to 172.22 ppm, after applying a 
quadratic fit over each data segment as shown in Fig. 6
the fully corrected data are shown in Fig. 7.
This is the
time series that will be used later to form power spectra
and search for evidence of p-modes.  The time series {\em rms}
here is 163.86 ppm.  The deadtime corrected count level
would imply a limit of 141.93 ppm at 30 seconds.
Based on these numbers the
removal of artifacts has provided data within 15\% of the Poisson limit.

\begin{figure}[h!]
\resizebox{\hsize}{!}{\includegraphics{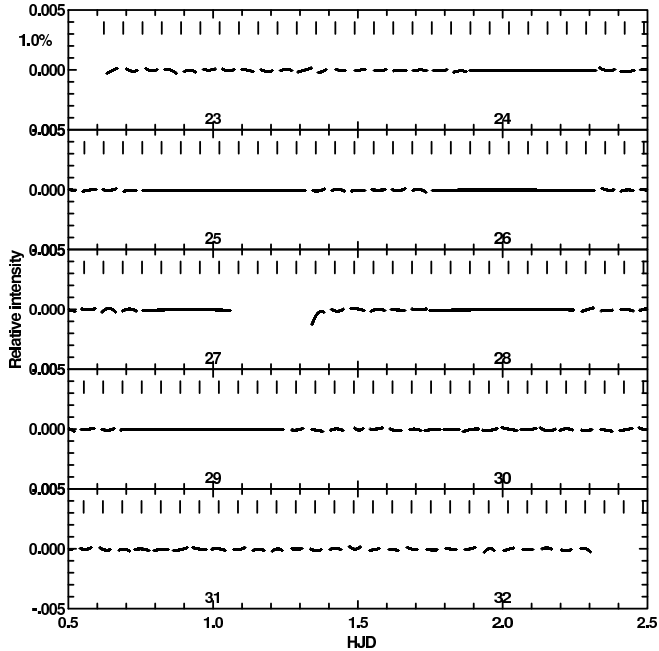}}
\caption{Final corrections derived after the iterative
solution and application for slow-drift, orbital waveform
and HV ramp changes.  The solution follows from a simple
quadratic polynomial fit to each data segment.
\label{fig6}}
\end{figure}

\begin{figure}[h!]
\resizebox{\hsize}{!}{\includegraphics{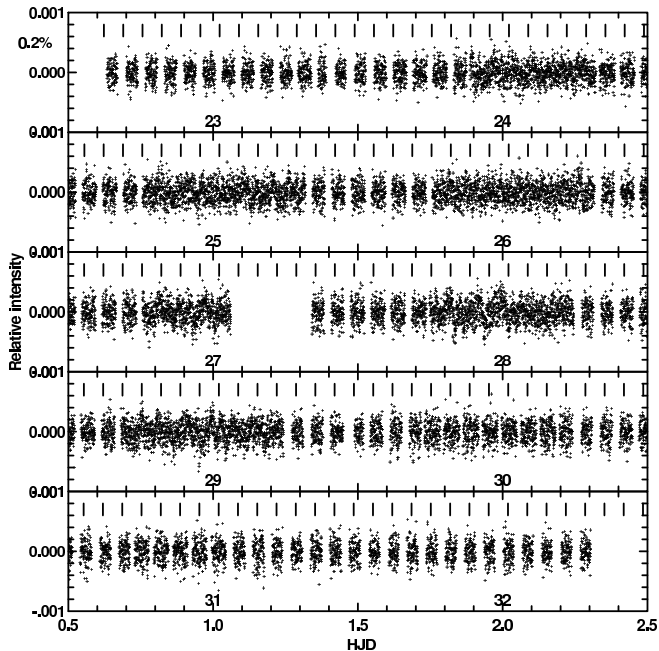}}
\caption{Same as Figure 1, but after having removed all
of the data artifacts discussed above.  Note that relative
to Figure 1 the vertical plot scale has changed by a 
factor of 5 to a full range of 0.2\%.  The rms of these 
30 second sums (ignoring 6 points deviant at more than
4-$\sigma$) is 163.86 ppm.
\label{fig7}}
\end{figure}

\subsection{Effect on Power Spectrum Frequency Content}

With the corrected time series in hand as shown in Fig. 7,
we are now ready to explore evidence for coherent 
oscillations.  With an rms of 163.86 ppm over 20202 data
points retained (dropping 6 with 4-$\sigma$ deviations),
the noise level is expected to be 1.15 ppm globally in
amplitude, likely lower and near 1.0 ppm at 
frequencies beyond 1.5 mHz.
A power spectrum is shown in
Fig. 8 over 0.5 to 4.0 mHz.  Below
about 1 mHz the reduction steps discussed above, coupled with large systematics
do not provide useful information on any stellar variations.
Above about 2.5 mHz inspection shows that the resulting power
spectrum seems only to reflect noise.  Based on published
scaling relations and knowledge of the stellar parameters,
we would expect oscillations peaking over 1.5--2.0\,mHz
at perhaps 30\% higher than solar amplitude, with a large
separation of some 87\,$\mu$Hz.
The errors on stellar parameters, in particular the mean
stellar density of 0.589\,g\,cm$^{-3}$, and -0.103, +0.066 1-$\sigma$ errors
from \citet{win09}, project to a large separation of 
87.3\,$\mu$Hz with a 1-$\sigma$ range of 79.3--92.0\,$\mu$Hz.

\begin{figure}[h!]
\resizebox{\hsize}{!}{\includegraphics{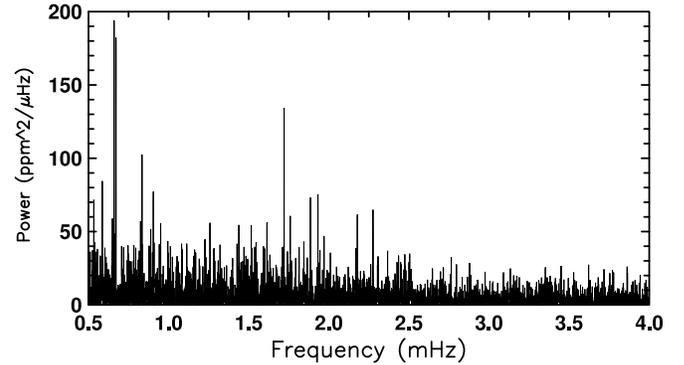}}
\caption{The power spectrum for the corrected data 
on HD 17156 as shown in Fig. 7.  The mean background
noise level near 4 ppm$^2/ \mu Hz$ is evident above 2.5 $\mu$Hz.
The spectrum below 1 mHz is likely still contaminated by systematics.
\label{fig8}}
\end{figure}

The upper panel of Fig. 9 shows a restricted range of 1.0
to 2.6 mHz for the power spectrum.
A window function is shown as the middle
panel in Fig. 9.  This was formed by generating a time
series on the observed cadence with very low amplitude
white noise and a large amplitude sinusoid at the same 
(1.7211 mHz) frequency as the highest peak in HD 17156.
The most prominent sidelobes are at multiples of the
{\em HST} orbit as expected, with smaller features 
reflecting the daily changes forced by SAA avoidance.

\begin{figure}[h!]
\resizebox{\hsize}{!}{\includegraphics{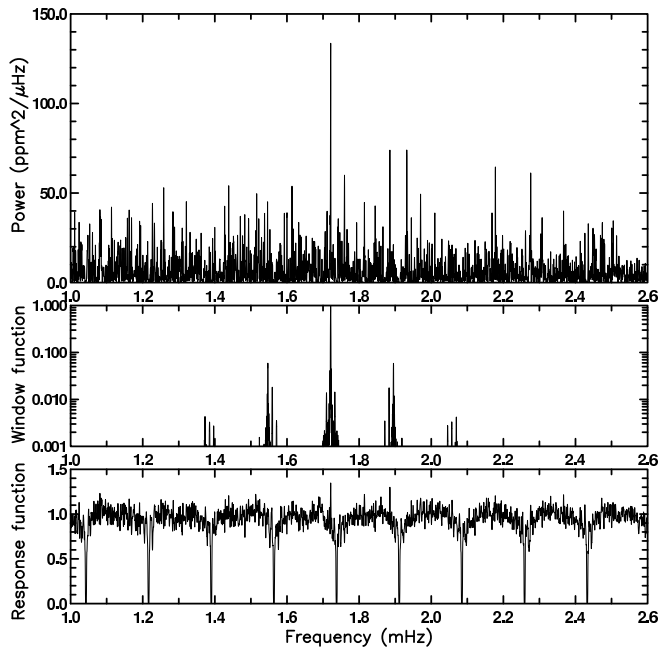}}
\caption{The upper panel shows the power spectrum 
over a restricted range of 1.0--2.6\,mHz.  The adopted frequency resolution of
0.25\,$\mu$Hz oversamples by about a factor of 5.
The middle panel shows the window function for the
power spectrum from injecting a strong sinusoid
at the times of observations
at the frequency of the strongest peak in the upper panel.
The lower panel shows
the response, or transfer function of the data reduction
procedure used in removing artifacts from these data.
\label{fig9}}
\end{figure}

Some of our reduction steps discussed above have been 
rather drastic.  In particular the orbital waveform 
correction shown in Fig. 4 will suppress any real 
oscillation frequencies that happen to coincide with
{\em HST} orbital harmonics (multiples of the 173.828\,$\mu$Hz
orbital frequency).  To quantify this we have injected
a test signal of 60 ppm into the raw data (Fig. 1),
then performed all of the reductions detailed in Figures
2 -- 6, following this with a new power spectrum.  
This is done at every frequency for which
the power spectrum is evaluated.
The ratio of power at the injection frequency to 
the input value then forms the response 
function shown in the bottom panel of Fig. 9.
As expected, signals at harmonics of the {\em HST} orbit 
are strongly suppressed.  However, the portion of 
frequency phase space in which signal power is
suppressed by more than 20\% (response function $<$0.8)
is only 4.7\%.  Knowledge of the response function 
should be taken into account in searching for evidence
of p-mode oscillations.
The response function is fixed in frequency, while of course
the sidelobes of the window function shift in concert with
source mode frequencies.

\section{EVIDENCE FOR STELLAR OSCILLATIONS}

The asteroseismic analysis of the HD 17156 data has been performed
using the pipeline developed at the Kepler Asteroseismic Science 
Operations Center as described in \citet{chr08}. A more extended discussion of 
Kepler pipeline analysis tools can be found in \citet{hub09}.
The outline of analysis steps follows: (1) Calculation of the power spectrum.
(2) Application of a matched filter (modified approach differing
from, but with similarities to the usual comb filter) method to
determine the large frequency
separation. (3) Calculate the folded power spectrum using the large 
frequency separation.  (4) Identify relation of $l$ mode 
frequencies relative to the large separation.
(5)  Derivation of individual frequencies and fit to
the asymptotic relation.  (6) Calculation of the oscillation amplitudes
corresponding to the radial, $l$ = 0 modes.

Oscillation frequencies for low-degree, p-modes are well approximated by
a regular series of peaks for which the oscillation frequencies are 
given by the approximate asymptotic relation:
\begin{equation}
\nu_{nl} \approx \Delta \nu_0 (n \, + \, l/2 \, + \, \epsilon) - D_0 l(l+1)
\end{equation}
where $\Delta \nu_0 = (2 \int ^R_0 dr/c)^{-1}$ corresponds to the
inverse of the sound travel time across the stellar diameter, and closely
relates to the stellar mean density via:
\begin{equation}
\Delta \nu_0 \approx 135 (M_* / R^3_*)^{1/2} \,\mu {\rm Hz}
\end{equation}
where $M_*$ and $R_*$ are the stellar mass and radius in solar units,
and the large separation for the Sun is approximately 135\,$\mu$Hz.

In the above equations $n$ is the radial order, and $l$ the angular degree of 
trapped oscillation modes.  $D_0$ is sensitive to the sound speed near the
stellar core, and $\epsilon$ is a correction factor absorbing minor frequency
dependent corrections sensitive to the stellar surface layers.
A recent review of the theory of solar-like oscillations may be found
in \citet{chr04}.

Although individual peaks may be visible in the power spectrum first
shown in Fig. 8, guided by an expectation given the scaling relations
of \citet{kje95} for largest amplitudes of about 7 ppm near 1.65 mHz,
these results are in a low-SNR regime in which reliance on existence of
evenly spaced peaks in frequency must be an inherent part of the 
process of coaxing information on the oscillations from the data.

A smoothed version of the power spectrum is illustrated in Fig. 10 which
clearly shows the excess of power for HD 17156.
The philosophy behind this analysis
is developed in order to avoid the stochastic nature of the excitation
and damping of individual oscillation modes. In order to measure the
oscillation amplitude in a way that is independent of these effects,
Kjeldsen et al. (2005, 2008) have suggested a method that involves
heavily smoothing the power spectrum in order to
produce a single hump of excess power that is insensitive to the
fact that the oscillation spectrum has discrete peaks.
Following Kjeldsen et al. (2008) we smoothed the power spectrum
by convolving (in power) with a Gaussian having a FWHM of 4 times 
the large separation.

\begin{figure}[h!]
\resizebox{\hsize}{!}{\includegraphics{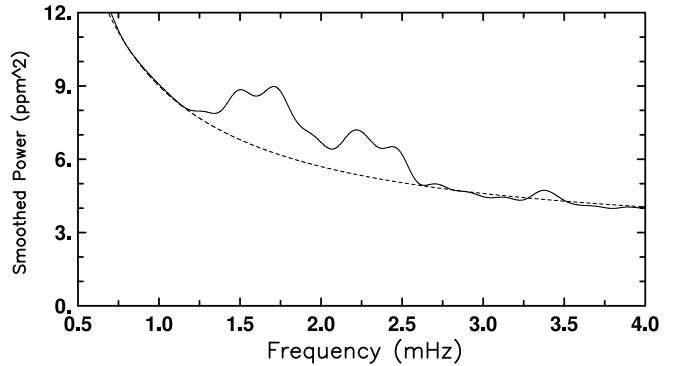}}
\caption{Smoothed power spectrum for HD 17156 shown as the
solid line with a fit to the background from frequencies over 0.5 to 1.2
and above 2.6 mHz shown as a dashed line.  The power excess of 
modes over 1.3 -- 2.4 mHz in the expected domain for oscillations
on HD 17156 is clear.
\label{fig10}}
\end{figure}

Following \citet{chr08} and \citet{hub09} we first determine 
the value of the large separation that best fits the power spectrum,
as was outlined by \citet{chr07}.
The idea is to match to the power spectrum a series of peaks described
such that they follow the asymptotic relation (Eq. (2)). 
To suppress the effect of the last term in that equation
(giving rise to the so-called small frequency separation; see below)
and avoid being sensitive to the small deviations from the asymptotic
description, the analysis is carried out for a smoothed power spectrum
$\bar P(\nu)$.
Specifically, this is obtained from the original power spectrum through
Gaussian smoothing with a FWHM of $3 \,\mu$Hz.
We now sum the power at uniformly spaced frequencies 
$\nu_k = \Delta \nu ( k/2 + \epsilon_0)$ corresponding to the
leading-order asymptotic expression, by calculating
\begin{equation}
\CF(\Delta \nu, \epsilon_0) = \sum_{k = 2 n_0 - \Delta k}^{2 n_0 - \Delta k}
\bar P(\nu_k) \; ,
\end{equation}
as a function of trial values of $\Delta \nu$ and $\epsilon_0$,
for a suitable central radial order $n_0$ and a suitable range $\Delta k$.
For each value of $\Delta \nu$ we determine the maximum 
$\CF_{\rm max}(\Delta \nu)$ of $\CF(\Delta \nu, \epsilon_0)$ as a function
of $\epsilon_0$.
This defines what we call {\it the matched filter response},
as a function of $\Delta \nu$.
Note that the procedure essentially determines the almost uniform separation
of $\Delta \nu/2$ between the nearly degenerate peaks corresponding to
even and odd degrees (this will be plotted transformed to $\Delta \nu$).
In our analysis we varied
$\epsilon_0$ between 1.0 and 1.5 (typical values found from stellar models),
and took $\Delta k = 5$, to include a total of 11 peaks in the analysis.
The result of using trial
values for the large separation between 40 and $120 \,\mu$Hz
is shown in Figure 11, for $n_0$ = 19.5, 20, 20.5, 21 and 21.5.
The maximum matched filter response is found at $83.60 \,\mu$Hz and 
clearly does not depend on $n_0$ within the range used.
Note that sampling the spectrum at a frequency separation of $\Delta \nu/2$,
and over a fixed range of radial orders, leads to a single peak
in the response, unlike other types of comb analysis.

\begin{figure}[h!]
\resizebox{\hsize}{!}{\includegraphics{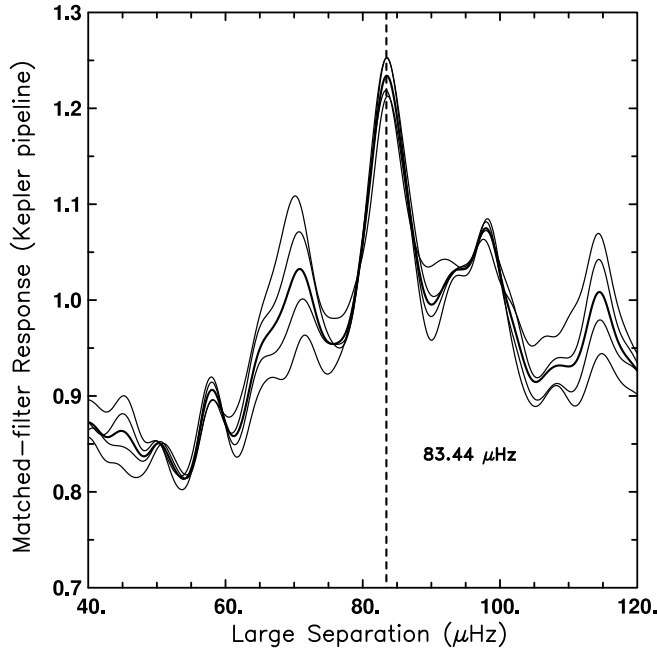}}
\caption{The matched filter-response function showing the large separation
$\Delta \nu_0$ for HD 17156.  The curves show results of 
invidividual foldings of the spectra using $n_0$ = 19.5, 20, . . . 21.5.
At 40 $\mu$Hz the search range is 860 $\pm$ 200 $\mu$Hz, while at
120 $\mu$Hz the range is 2580 $\pm$ 600 $\mu$Hz.
The more precisely determined value
from fits of individual frequencies to the 
asymptotic relation is shown at 83.44\,$\mu$Hz with the dashed line.
\label{fig11}}
\end{figure}

The results shown in Fig. 11 yielding $\Delta \nu_0 \approx 83.6\, \mu$Hz
are stable and robust. 
In particular, the same value of the large 
separation can also be found by considering only the first, middle or
last half of the data sets, and it also continues to be seen if the strongest
peak in the power spectrum at 1721.2\,$\mu$Hz is arbitrarily set to zero.  
Thus the 9.7 days of {\em HST} observations of HD 17156
with FGS2r have provided a secure detection of solar-like p-modes.
Further quantification and use of the large separation follows below.

The next stage in the analysis attempts to fix further details
of the oscillation frequencies.  A goal here is to determine
the small separation, which would provide constraints on the stellar age
if possible, and to provide individual frequencies for as many
modes as possible.
A necessary associated goal will be to determine mode identifications,
i.e. corresponding $n$ and $l$ with specific frequencies.

Fig. 12 shows the result of folding the amplitude spectrum at 83.5\,$\mu$Hz
(taking into account the 83.44 $\mu$Hz splitting from fits to individual
mode frequencies below)
over the range of 1.2--2.4\,mHz.
The presence of strong contrast in this figure is simply another way 
of demonstrating that 83.5\,$\mu$Hz is the correct large separation,
e.g. folds at other values would show less contrast similar to the fall-off 
in distribution of the matched-filter response function of Fig.~11
away from 83.5\,$\mu$Hz.
The expectation is that modes of $l = 0$, 1, and 2 only will 
be visible, with the modes at $l = 0$, and 2 nearly degenerate
except for the small separation term of Eq.~(2).
This leads to the expectation that a single isolated peak in 
this figure will correspond to $l = 1$, while a doubled peak
with frequencies differing by much less than the large separation
will represent the $l = 0$ and 2 modes.  Guidance from theoretical
models to be further discussed in \S 5 has led to the identification
of modes provided in the caption to Fig. 12 and in Table~1.

\begin{figure}[h!]
\resizebox{\hsize}{!}{\includegraphics{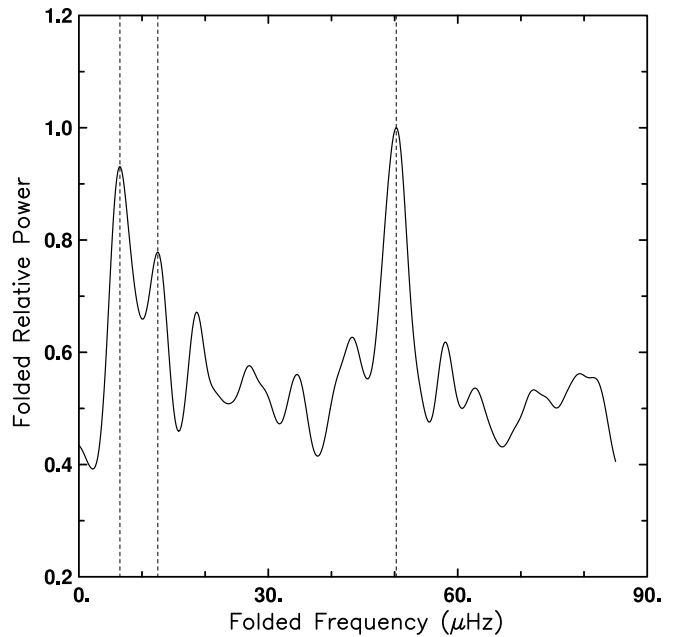}}
\caption{The power spectrum of Fig. 9 after smoothing by 3\,$\mu$Hz
(FWHM) and averaging over
successive slices of 83.5\,$\mu$Hz over the domain of 1.2--2.4\,mHz.
The dashed lines indicate features which left to right correspond
to modes of $l$ = 2, 0, and 1 respectively.
\label{fig12}}
\end{figure}

One might also expect to have a higher signal from the $l = 0$
modes than from the $l = 2$ modes in Fig. 12 which is not the case.
Kjeldsen, Bedding \& Christensen-Dalsgaard (2008) give amplitude
ratios of $l = 2/0$ for the Sun as 0.81, 0.75, and 0.67 at
wavelengths of 402, 500 and 862 nm respectively, which for the flux
weighted centroid of these FGS observations at about 550 nm
implies an expected ratio of 0.74.  It might also be expected
that the number of $l = 0$ modes detected would exceed $l = 2$;
again this is not the case. In fact, due to the stochastic
nature of the excitation it would not be surprising for modes
with $l = 2$ sometimes to exceed the $l = 0$ mean amplitudes.
Of the three stars with high S/N oscillations discussed in
Chaplin et al. (2010), KIC 3656476 has a $\Delta \nu$ nearest
HD 17156 and it shows equal numbers of detected $l = 0$ and 2 modes,
and in 2 of the 4 jointly highest pairs of these the $l = 2$ amplitudes
are larger than $l = 0$.  

Table 1 provides the individual frequencies that follow from 
identifying the
10 highest peaks in the power spectrum. 
These all have a signal-to-noise ratio in excess of 4 in
the power spectrum of Fig. 9. From the asymptotic relation we identify 8
of the 10 highest frequencies. Their degree and order are shown in Table 1.
The errors in Table 1 are estimated from simulations of the time series.
We created a large number of stochastically excited modes with a mode lifetime
of 3 days (similar to the lifetime of modes in the Sun) and estimated the
accuracy with which we could detect the frequencies
(taking peak amplitudes at same SNR as in the present data)
in a time series with same sampling as the present data.
A lifetime longer than 3 days would have resulted in smaller estimated
frequency errors than indicated,
while the errors would be larger if the mode lifetime is shorter than 3 days.
The identification of the 8 frequencies that fit the comb 
structure should not be seen as an unbiased frequency
identification since it relies on selection of the modes
that fit the comb structure for l=0, 1 and 2.
We also assume the existence of the asymptotic relation 
and the structure of stellar model frequencies to fix the
identification. Another possible approach would be to use
a wider frequency range when identifying the modes
or use modes with lower amplitude. We have tried to
use more modes and the fit to the stellar models does not
depend (within the error bars) on the exact number of peaks
included in the analysis. In the end we decided to use the
10 strongest peaks as a way to ensure that we only work
with the most significant peaks and of those 10, 8 peaks
agreed with the asymptotic relation corresponding to the 
comb power in Figure 12.

We have carried out a weighted least-squares fit 
of the asymptotic relation, Eq. (2), to the 8 identified frequencies,
estimating the errors in coefficients from a Monte Carlo simulation.
The resulting coefficients are:
$\Delta \nu_0 = 83.44 \pm 0.15 \,\mu$Hz, $D_0 = 0.90 \pm 0.19 \,\mu$Hz,
and the surface term $\epsilon = 1.15 \pm 0.04$.

Figure 13 shows the power spectrum from Fig. 9 after smoothing
with a Gaussian of FWHM = 3 $\mu$Hz on which the individual mode
frequencies of Table 1 are included.
Over the frequency range 1.2 -- 2.5 mHz 8 of the 10 highest peaks
are flagged as identified.

\begin{figure}[h!]
\resizebox{\hsize}{!}{\includegraphics{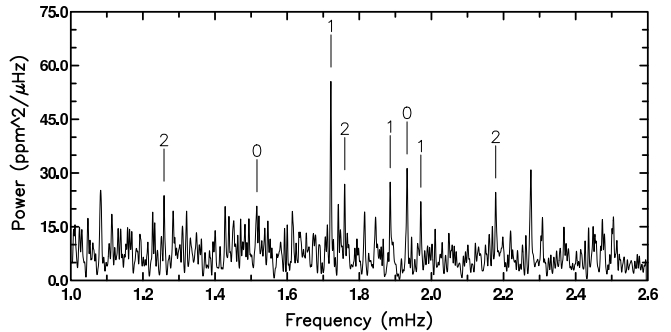}}
\caption{A version of power spectrum as in Fig. 9 after a Gaussian
smoothing of 3 $\mu$Hz FWHM has been applied.
Frequencies of individual modes as listed in Table 1 are indicated
with vertical bars at the listed frequency.  The numerical label
of 0, 1, or 2 provides the $l$ value.
\label{fig13}}
\end{figure}

The final step in direct analyses of the amplitude spectrum is to 
assess the amplitude per mode from a smoothed power spectrum, and 
distribution of amplitudes with frequency following the method
described in \citet{kje08}, and \citet{hub09}.
Converting the smoothed power
spectrum (Fig. 10) to power density by multiplying by the effective length of 
the observing run followed by fitting and subtracting
the background noise and then multiplying by the large separation divided 
by the number of p-modes peaks per radial order scaled to the sensitivity 
of radial modes we may calculate the mean power per radial mode.
The square root is then taken in order to convert to amplitude per
oscillation mode (radial modes).
The peak amplitude
is about 7 ppm for $l$ = 0, quite consistent with pre-observation
estimates made using the published stellar parameters and \citet{kje95}.
The distribution of amplitudes
for radial modes
in HD 17156 is contrasted to those in the Sun in Figure 14.
The extended, non-Gaussian
distribution of amplitudes, albeit not well determined here, is similar
to those found for Procyon \citep{are08}.

\begin{figure}[h!]
\resizebox{\hsize}{!}{\includegraphics{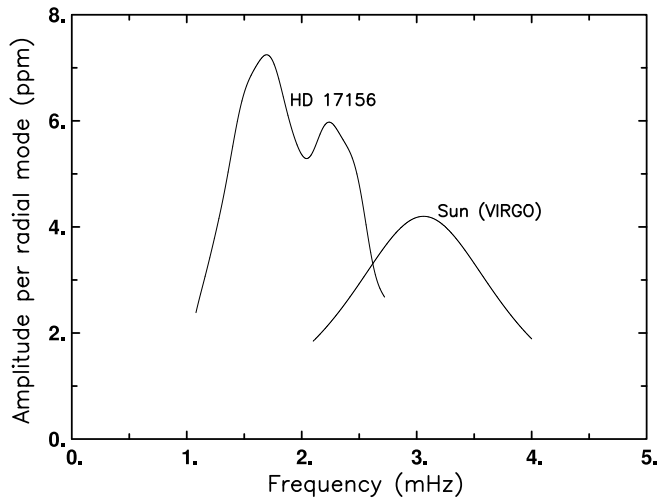}}
\caption{The amplitude per radial mode distribution for HD 17156 
contrasted to that for the Sun.  See text for discussion.
\label{fig14}}
\end{figure}

\section{STELLAR EVOLUTION MODELS AND INTERPRETATIONS}

\subsection{ASTEC -- ADIPLS Analyses}

The stellar evolution models are based on ASTEC \citep{chr08a} and the 
associated eigenfrequency analysis code ADIPLS \citep{chr08b}.
Briefly, the models use the OPAL equation of state \citep{Rogers1996}
and opacities \citep{Iglesi1996}, and the NACRE nuclear parameters
\citep{Angulo1999}.
The temperature gradient in the convection zone was computed using
the \citet{Bohm1958} mixing-length treatment,
with a mixing-length of $\alpha_{\rm ML} = 2.00$ pressure scale heights,
roughly calibrated to the corresponding solar models.
In some cases convective core overshoot was included,
over a distance of $\alpha_{\rm ov} \min(r_{\rm cc}, H_p)$,
where $r_{\rm cc}$ is the radius of
the convectively unstable region and
$H_p$ is the pressure scale height at this point;
the overshoot region was assumed to be fully mixed and
adiabatically stratified.
Diffusion and settling of helium were treated according to the
simplified formulation of \citet{Michau1993}.
It was assumed that the initial abundances $X$ and $Z$ by mass 
of hydrogen and helium were related by $X = 0.77 - 3 Z$, 
corresponding to a galactic enrichment $\Delta Y$ of helium determined
by $\Delta Y = 2 Z$; 
the values of $X$ and $Z$ were characterized by the observed
[Fe/H], assuming a present solar surface composition with $Z/X = 0.0253$.

\begin{figure}[h!]
\resizebox{\hsize}{!}{\includegraphics{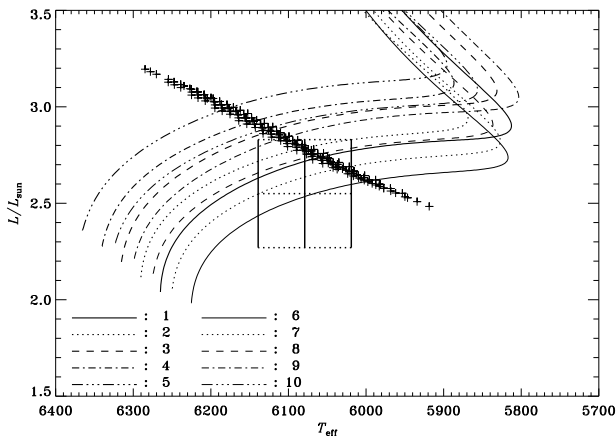}}
\caption{Theoretical HR diagram with selected evolutionary tracks,
corresponding to the models defined in Table 2. 
The '+' indicate the models
along the full set of evolutionary sequences minimizing the
difference between the computed and observed frequencies.
The box is centered on the
$L$ and $T_{\rm eff}$ as given in \citet{win09},
with a size matching the errors on these quantities
($T_{\rm eff}$ error from Table 3, $L$ from \citealp{win09}).
\label{fig15}}
\end{figure}

Evolution tracks have been computed to match the classical
observed parameters as given in \citet{fis07} and \citet{amm09},
leading to $T_{\rm eff} = 6082 \pm 60 \, {\rm K}$ and
${\rm [Fe/H] = 0.24 \pm 0.03}$.   
A grid of models was computed,
varying the mass between 1.26 and $1.33 \Msun$ in steps of 0.01 $\Msun$.
with composition, characterized by $Z$,
corresponding to [Fe/H] varying between 0.18 and 0.30 in steps of 0.02,
and with $\alpha_{\rm ov} = 0$, 0.05, and 0.1.
All models included diffusion and settling of helium.  
Figure~\ref{fig15} shows the theoretical HR diagram with selected 
evolutionary tracks.
The associated Table 2 lists the models along the evolutionary
tracks at which the computed eigenfrequencies best match 
the observed frequencies (see \S 5.2).

\subsection{Fitting the Observed Quantities}

We have carried out least-squares fits of the observed frequencies 
in Table~1 and
the observed effective temperature to the computed grid of models.
For each model in the grid we computed the mean square difference 
of the model frequencies $\nu_{nl}^{\rm (mod)}$ from
the observed frequencies $\nu_{nl}^{\rm (obs)}$:
\begin{equation}
\chi_\nu^2 = {1 \over N - 1} 
\sum_{nl} \left( {\nu_{nl}^{\rm (obs)} - \nu_{nl}^{\rm (mod)}
\over \sigma(\nu_{nl})} \right)^2 \; ,
\label{eq:chisqnu}
\end{equation}
where $N = 8$ is the number of observed frequencies
and $\sigma(\nu_{nl})$ is the estimated error in the frequencies.
This was minimized along each evolution track, characterized by
the parameters $\{M, Z, \alpha_{\rm ov}\}$.
We first determined the model, $\CM_{\rm min}$,
in the evolution sequence with the smallest value of $\chi_\nu^2$.
We then assumed that the best-fitting model for these parameters
could be obtained from the frequencies $\nu_{nl}^{\rm (mod)} ( \CM_{\rm min})$
by scaling,
\begin{equation}
\nu_{nl}^{\rm (mod)} = r \nu_{nl}^{\rm (mod)} ( \CM_{\rm min}) \; ,
\end{equation}
and determined $r$ by minimizing $\chi_\nu^2$.
The results presented in the following are based on these
resulting minimal $\chi_\nu^2$ along the evolution tracks.
According to Eq.~(3) the value of $r$ so obtained
determines the radius of the best-fitting model
as $R = r^{-2/3} R( \CM_{\rm min})$;
the remaining model quantities, including the effective temperature
$T_{\rm eff}^{\rm (mod)}$,
were then determined by linear interpolation in radius to this value.
Finally, the departure from the observed effective temperature
$T_{\rm eff}^{\rm (obs)}$ was included in the final 
\begin{equation}
\chi^2 = \chi_\nu^2 + \left( {T_{\rm eff}^{\rm (obs)} - T_{\rm eff}^{\rm (mod)}
\over \sigma(T_{\rm eff}) } \right)^2 \; ,
\label{eq:chisq}
\end{equation}
where $\sigma(T_{\rm eff})$ is the standard error on the effective temperature.

\begin{figure}[h!]
\resizebox{\hsize}{!}{\includegraphics{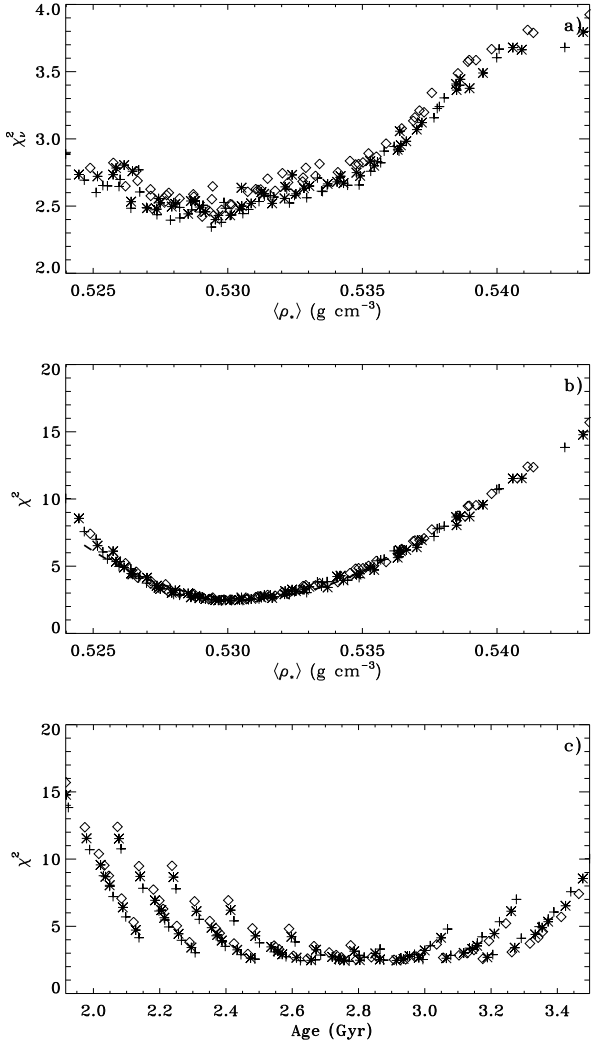}}
\caption{Results of fitting the observed frequencies and effective
temperature to the grid of stellar models (see text for details).
Plusses, stars and diamonds correspond to models with $\alpha_{\rm ov} = 0$
(no overshoot), 0.05, and 0.1, respectively.
Panel (a) shows the minimum mean square deviation $\chi_\nu^2$ 
of the frequencies (cf.\ Eq.~\ref{eq:chisqnu}) along each evolution track,
against the mean density $\rhomean$ of the corresponding models.
Panel (b) similarly shows the combined $\chi^2$ (cf.\ Eq.~\ref{eq:chisq});
here the dashed curve is a parabolic fit to those points that have
$\chi^2 \le 7.2$ (see text).
Finally, panel (c) shows $\chi^2$ against the age for the models that
minimize $\chi_\nu^2$;
the different ridges correspond to the different masses in the grid, 
the more massive models resulting in a lower estimate of the age.
\label{fig:chisq}}
\end{figure}

The results of these fits are illustrated in Fig.~\ref{fig:chisq}.
A key goal of this analysis is to determine the mean stellar density 
$\rhomean$ of the star; thus panels a) and b) show $\chi_\nu^2$ and
$\chi^2$ against $\rhomean$ for all the models in the grid.
It is evident that minimizing $\chi_\nu^2$ along the evolution tracks
leads to a narrow range of mean density, with well-defined minima 
at fixed $\alpha_{\rm ov}$.
It is interesting that some preference is found for models without
overshoot, although the difference between the three cases is
of limited significance.
Including also the constraint of the observed $T_{\rm eff}$ (panel b) 
produces a well-defined minimum in $\chi^2$, shifted towards slightly
higher $\rhomean$.
To quantify the location and width of this minimum we have fitted a parabola
to those points that have $\chi^2 \le 3 \chi_{\rm min}^2$,
where $\chi_{\rm min}^2 = 2.40$ is the minimum over all the sequences in the
grid.
This parabola has a minimum at $\rhomean = 0.5301 \, {\rm g \, cm^{-3}}$,
and indicates a standard error in $\rhomean$ of $0.0031 \, {\rm g \, cm^{-3}}$.
Augmenting that by a factor of $\sqrt{2}$ to account for further possible
systematic errors we arrive at our final estimate:
\begin{equation}
\rhomean = 0.5301 \pm 0.0044 \, {\rm g \, cm^{-3}} \; .
\end{equation}

The error estimate in Eq. (8) is derived from the curvature of 
the $\chi^2$ surface.  This estimate may also be cast in the 
form of a confidence interval in the manner described by, 
e.g., \citet{kal10}.  This is done by computing
the relative probability $\exp^{-\chi^2/2}$ for each model, 
and then normalizing so that the sum of the probabilities over 
all models is unity.  The desired confidence interval is then 
obtained by forming the marginal distribution of these probabilities
with respect to the stellar mean density, and determining
the range of densities that contain the desired fraction of the 
total probability.  In the current case, $\chi^2$ as a function 
of $<\rho_*>$ is very well approximated by a parabola;  as a result,
the marginal probability distribution is almost indistinguishable
from a Gaussian, and the 68\% confidence interval agrees closely
with the standard error in $<\rho_*>$ of 0.0031 g cm$^{-3}$ given above.

As illustrated in Fig.~\ref{fig:chisq}c the fit also provides a constraint
on the stellar age, although substantially affected by the spread in the
mass in the grid, producing a broad nearly flat minimum in $\chi^2$.
On the basis of the plot we estimate the age as $2.8 \pm 0.6$\,Gyr.

In Fig. \ref{fig15} we have indicated the models minimizing $\chi_\nu^2$
for all the parameter sets in the grid.
These clearly fall in a tightly confined region in the diagram, corresponding
to the strongly constrained mean density.
In order further to illustrate the properties of the fit we have considered
models with masses between $1.28$ and $1.32 \Msun$ and $Z = 0.0299$
([Fe/H]$ = 0.24)$ and $0.0338$ ([Fe/H]$ = 0.30)$,
the latter case including the model minimizing $\chi_\nu^2$.
For each pair $(M, Z)$ we selected the sequence leading to the smallest
$\chi_\nu^2$.
These are the evolution tracks plotted in Fig.~\ref{fig15} and
with properties listed in Table~2.

\begin{figure}[h!]
\resizebox{\hsize}{!}{\includegraphics{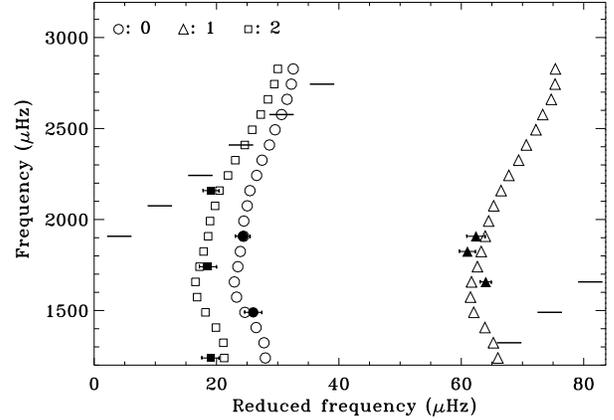}}
\caption{
The asteroseismic \'echelle diagram comparing the observed
frequencies of Table~1 (filled symbols with $1 \sigma$ error bars)
with values from the theoretical
model minimizing $\chi_\nu^2$ (cf.\ Eq.~\ref{eq:chisqnu}; open symbols);
the model (model no 8 in Table 2)
has a mass $M = 1.30 \Msun$, ${\rm [Fe/H]} = 0.30$ and an age of 2.67\,Gyr.
The horizontal axis shows frequency distribution within successive
83.6\,$\mu$Hz slices, starting at a frequency of
68.8\,$\mu$Hz, while the vertical axis shows the
starting frequencies of the slices.
The short horizontal lines indicate the frequency intervals
where the response function drops below 50\% due to orbital filtering
(cf.\ Fig.~\ref{fig9}).
\label{fig:echelle}}
\end{figure}

To illustrate the quality of the fit of the computed frequencies to the 
observations, Fig.~\ref{fig:echelle} compares the observed and 
computed frequencies in an {\it \'echelle diagram}
\citep[][see caption]{Grec1983}, for the model in the grid which minimizes
$\chi_\nu^2$.
It is evident that the fit is excellent.

As an alternative fit to the model grid, which is independent of the 
individual frequencies, we have selected, on each evolution track, the model
that matches the large separation $\Delta \nu$.
Averaging the resulting mean densities over those models where $T_{\rm eff}$
differs by less than $2 \sigma$ from the observed value yields
$\rhomean = 0.5290 \pm 0.0030 {\, \rm g \, cm^{-3}}$, 
fully consistent with the more detailed fit.

We finally note that a potential problem in fitting solar-like oscillations
is the effect of the near-surface layers on the frequencies which is not
properly taken into account in the adiabatic modeling considered here.
This effect is well known in analysis of helioseismic data
\citep[e.g.,][]{chr96}, 
where it can be isolated, owing to the availability of modes over a broad 
range of degree.
\citet{kje08} suggested a procedure to eliminate the effect in the
analysis of asteroseismic data by assuming a similar functional form as in
the solar case, but a potentially different amplitude, to be determined
as part of the fit; 
in addition, the procedure results in an estimate of the mean density
of the star, through scaling from a suitable reference model.
We have applied this to the observed frequencies in Table 1;
this resulted in insignificant changes to the obtained fits and
stellar parameters.

\subsection{Comparing Asteroseismic and Planet-transit Mean Densities with 
Surface Gravity}

Asteroseismology, and transit light curve modeling when very precise
relative photometry is available, can both provide very precise
(even accurate) determinations of the mean stellar density of stars.
Since density is a simple function of stellar mass and radius, both
asteroseismology and transit light curve modeling tightly constrain
allowed choices of mass and radius in individual cases.
However, arriving at unique estimates of the stellar mass and radius
separately, as for example are needed if one wishes to use these
to provide optimal estimates of the mass and radius of hosted planets
as needed to advance and challenge the theoretical study of 
extrasolar planet formation and evolution, requires independent input.

\begin{figure}[h!]
\resizebox{\hsize}{!}{\includegraphics{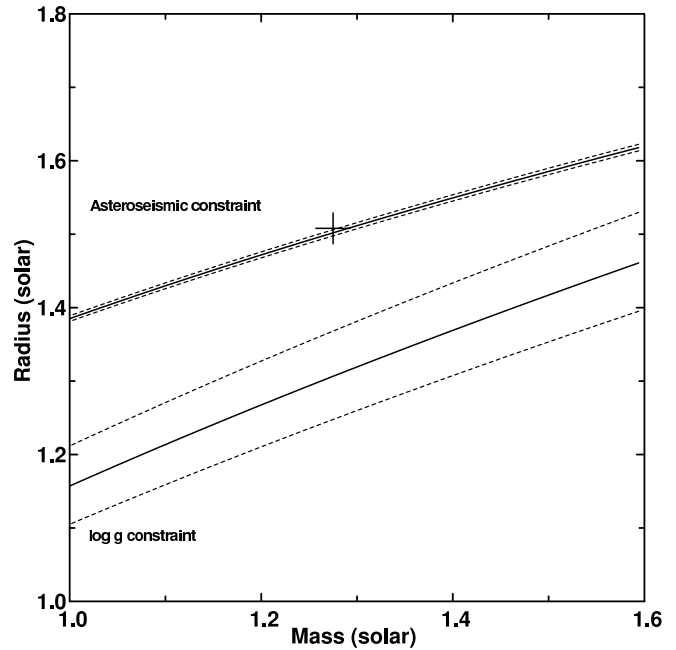}}
\caption{The upper curves show the allowed $M_* , R_*$ values that
satisfy the asteroseismology large separation constraint (solid curve
prefered value of $\rhomean = 0.5301\,{\rm g \, cm^{-3}}$,
dashed lines the 0.0044 error
offsets from this).  The lower curves show the corresponding 
values that satisfy the spectroscopic $\log(g)$ value of 4.31, with solid
and dashed curves preferred and 0.04 error offsets.  The large plus
symbol shows the preferred value of $M_* , R_*$, and allowed error 
range resulting from the transit light curves analyses \citep{nut09} --
note the excellent consistency with asteroseismology.
\label{fig:grav}}
\end{figure}

For both asteroseismology and transit light curve modeling the
classical approach to fixing the stellar mass and radius independently,
given a measured stellar density, relies on stellar evolution models
that best match observed constraints, typically
some combination of temperature, metallicity and surface gravity
(luminosity in our case with the known parallax, although we use luminosity
only as a consistency check),
hence bringing in inferences from spectroscopy.
Accurate estimates of stellar masses and radii then depend
on correct models of both stellar evolution and stellar
atmospheres.  
How good are these?
Figure~\ref{fig:grav} illustrates one outstanding issue, that may perhaps exist
primarily for stars in the high metallicity domain, as is the case
for HD 17156, and for many exoplanet hosts given the strong correlation
of high metallicity and existence of such planets (\citealp{fis05}; \citealp{val05}).
Spectroscopic studies have fixed $\log(g)$ to be $4.29 \pm 0.06$ \citep{fis07}
and $4.33 \pm 0.05$ \citep{amm09}, with very consistent determinations
of $T_{\rm eff}$ and [Fe/H] between the two independent studies as well;
see Table 3 for summaries.
In Fig.~\ref{fig:grav} we show the range of stellar masses and radii
that jointly
meet either the asteroseismic constraint for HD 17156 where $M/R^3$ is
a constant, or the constraint imposed by published spectroscopic $\log(g)$
values which requires $M/R^2$ to be a constant.  In this diagram an 
ideal result would be that the two sets of joint $M, R$ constraints
cross at the location of the preferred values obtained from other
considerations.  Instead, we are left with the result that there is 
no agreement, the asteroseismic result for radius 
remains nearly 3 $\sigma$ away from the $\log(g)$ constraint near the
preferred mass.  Note that the transit light curve analyses of 
\citet{nut09} also impose a constraint with the functional form
of that from asteroseismology, in relative terms this would have
a central value that falls near, but slightly above the preferred
asteroseismic value, with error offsets about 5 times larger
than those shown by asteroseismology.  Clearly the spectroscopic
$\log(g)$ of 4.31 is well above the asteroseismic (or transit light curve)
value of about 4.19 for HD 17156.
Perhaps in the high metallicity domain of interest for HD 17156,
and many exoplanet hosts, the stronger lines result in biases
for $\log(g)$.

\subsection{Calibration of Stellar Mass and Radius to Eclipsing Binaries}

Figure~\ref{fig:grav} shows that radius estimates based on spectroscopic 
$\log(g)$ determinations
are problematic at the 10\% level, whereas estimates based
on transit timing and on asteroseismology are mutually consistent.
This consistency provides a necessary, but not sufficient condition
for using the stellar mean density as measured from transit light 
curves or from asteroseismology to provide the truly desired 
stellar masses and radii.
To verify that we can combine a mean density measurement with stellar
evolution models to obtain accurate masses and radii, we must test
our techniques against stars for which there are accurate and independent
measures of mass and radius.
Eclipsing binaries (EBs) comprise the only large set of stars for which such
information exists.
The mass and radius estimates derived from light curve and radial velocity
measurements of EBs are almost independent of theory, and
in many cases accuracies of better than 3\% can be obtained for both
quantities.
\citet{tor09} have recently published the fundamental data for all 
known EBs meeting this standard of accuracy, a list containing 190 stars.
\citet{bro09} applied the mean-density analysis to 156 of these
(all with masses $M_*$ satisfying $0.4 M_{\sun} \leq M_* \leq 5M_{\sun}$),
matching Yonsei-Yale (YY) models \citep{yi01} to observational estimates of
mean density, effective temperature, and metallicity.
Comparing radii and masses from model-matching to those measured directly
showed that matching the YY models generally yielded accurate results,
with systematic errors estimated to be smaller than 2\% in radius and
6\% in mass.
Larger errors can occur for cool and rapidly-rotating stars, in which
surface magnetic activity is thought to interfere with convective energy
transport, leading to radii that are larger than the YY models predict.
HD 17156 is, however, a slow rotator
that is somewhat hotter than the Sun.
For this star, systematic errors resulting from activity should
be small, certainly less than 1\% based on the expected rotation period
of 19 days mentioned in \S 2, and the lack of any photometric variations 
in excess of $\sim$0.1\% intrinsic to the star in our FGS photometry.

\subsection{Stellar Parameters from the Inferred Mean Density}

To obtain an independent estimate of the values of the basic stellar
parameters and their errors we have carried out a
Markov Chain Monte Carlo (MCMC) fit of stellar models to the observed
$T_{\rm eff}$ and [Fe/H] and the asteroseismically inferred 
$\rhomean$ \cite[see][for details]{bro09}.
The models were obtained from the Yonsei-Yale (YY) compilation \citep{yi01}.
As for the ASTEC models used in the asteroseismic analysis the YY models
use OPAL opacities and equation of state and include diffusion and settling
of helium, although with a somewhat different formulation.
Core overshoot is included, with a step-function dependence of overshoot
distance on stellar mass \citep[see][for details]{Demarq2004},
such that for the models relevant to the present fits $\alpha_{\rm ov}$
is probably generally equal to 0.1.

The inferred parameters are listed in Table 3. The mass, radius and luminosity
are very close to the values obtained from the asteroseismic analysis
(see Table~2).
The age inferred from the MCMC analysis, $3.2 \pm 0.3$\,Gyr,
is formally consistent with the asteroseismic value.
However, the difference of 0.4\,Gyr probably reflects systematic
differences between the ASTEC and YY evolution codes.
This deserves further investigation.

\section{DISCUSSION}

\subsection{Summary and Comparison with Previous Results}

We reported the detection of the asteroseismic large separation 
for HD 17156, and determined the mean stellar density resulting
from this to be $0.5301 \pm 0.0044$\,g\,cm$^{-3}$.  
The best determination before these observations comes from
\citet{win09} who relied on ground-based transit light curve
analysis to fix $\rhomean$ at the 1 $\sigma$ range 
of 0.486 -- 0.655\,g\,cm$^{-3}$, 
with a preferred value of 0.589\,g\,cm$^{-3}$.
Our density is significantly 
different than the \citet{win09} value, but well within their
original confidence range.

\subsection{Comparison with Joint {\em HST} FGS Transit Analysis}

Of more relevance to comparing our asteroseismic results to 
those from transit light curve based studies are the results
from independent {\em HST} FGS observations obtained as a 
part of this program.  \citet{nut09} find a mean stellar density
of 0.522$^{+0.021}_{-0.018}$ g cm$^{-3}$.  This transit-based
determination is 1.8 $\sigma$ from the asteroseismic result, while
the latter is only 0.4 $\sigma$ off from the transit analysis 
based on its larger error allowances.
Clearly of the two extreme outcomes possible from this first
comparison of the two independent techniques that both 
yield what may be referred to as direct determinations of 
$\rhomean$ we are firmly in the domain of mutual confirmation,
rather than a potential domain of disagreement calling for
further understanding of one or both techniques.  As noted in \S 5.3, however,
the same cannot be said for consistency with the spectroscopically
determined $\log(g)$ where there is a nearly 3 $\sigma$ discrepancy
in the sense that the spectroscopic $\log(g)$ is too large.

\subsection{Future Prospects from {\em Kepler}}

In the near future the {\em Kepler} Mission may be expected to
provide similar results in which both asteroseismology and 
transit light curve analysis will provide simultaneous constraints
for the mean stellar density of planet host stars.
The three previously known exoplanets in the {\em Kepler} field
of view will be observed at short cadence (58.8 seconds) throughout
the mission, thus supporting asteroseismology on these targets.
``Only" 512 targets may be followed at short cadence at any time,
the bulk of observations for over 150,000 stars with {\em Kepler}
will use the long cadence of 29.4 minutes which suffices for 
detection of planets via transits.  The {\em Kepler} throughput
is a factor of about 5.9 higher than that for the {\em HST} FGS2r
used for the observations in this paper, following from the use
of back-side illuminated CCDs on {\em Kepler}, and a broad 
bandpass of roughly 420 -- 880 nm, {\em despite the factor of
6.4 relative advantage in aperture area of HST compared
to Kepler}.  Furthermore, {\em Kepler} should reach a duty
cycle of nearly 100\% during the month-long observing blocks 
between short breaks for telemetering accumulated data to the
ground, compared to the uniquely high 72.6\% duty cycle reached
with {\em HST} for this observation taking advantage of 
a CVZ passage.  This results in a net advantage in terms of 
Poisson statistics limit of 2.27 magnitudes for {\em Kepler}
observations compared to those in this paper.  What our 
{\em HST} observations have provided for a $V$ = 8.17 star
in 10 days, should be possible with {\em Kepler} for a 
$V$ = 10.44 magnitude star in the same length of time.
Also weighing in favor of {\em Kepler} asteroseismology is 
an expected window function without the sidelobes resulting
from the orbit of {\em HST} and the daily passages
through the South Atlantic Anomaly.  And of course it should
be routine for {\em Kepler} to devote much longer observing 
periods to targets than was the case for this unusually long
{\em HST} observation.  The initial target catalog for {\em Kepler}
long cadence observations contains about 2500 targets brighter
than the level which should return Poisson limited precisions
per unit time as good or better than those discussed in this paper.
To reach the same S/N on expected oscillations in the three
previously known exoplanet hosts within the {\em Kepler}
field of view, as for these {\em HST} observations of HD 17156
should take about 2 months for TrES-2 at $V$ = 11.4, 8 days 
for HAT-P-7 at $V$ = 10.5, and 20 days for HAT-P-11 at $V$ = 9.6
taking into account the expected oscillation properties of each.





\acknowledgments

Coming at a time of significant stress on the {\em HST} project
these observations required the expert assistance of many 
individuals to develop and execute.  Especially noteworthy were
the skill and tireless efforts provided by Merle Reinhart
at STScI in expertly crafting the Phase II program to use all
available observing time in the orbits allocated, and the efforts
from Mike Wenz at Goddard Space Flight Center for shepherding the
proposal through pre-flight reviews and monitoring engineering 
performance during execution.  We thank the STScI Director, Matt 
Mountain, for the generous DD time award that made these results
possible. 
We thank Matt Holman and Jeff Valenti for discussion.
Financial support for this work was provided through program
GO/DD-11945 from STScI.



{\it Facilities:} \facility{HST (FGS)}.

\clearpage



\clearpage

\begin{table}
\begin{center}
\caption{Individual frequencies ($\mu$Hz) identified for HD 17156.\label{tbl-1}}
\begin{tabular}{cccc}
\tableline\tableline
$n$ & $l=0$ & $l=1$ & $l=2$ \\
\tableline
 13 &   ----- &  ----- &  $1258.2 \pm 1.4$ \\
 17 &   $1516.0 \pm 1.4$ &  ----- &  ----- \\
 19 &   ----- &  $1721.2 \pm 0.9$ &  $1759.4 \pm 1.4$ \\
 21 &   ----- &  $1885.4 \pm 1.3$ &  ----- \\
 22 &   $1932.3 \pm 1.2$ &  $1970.4 \pm 1.5$ &  ----- \\
 24 &   ----- &  ----- &  $2177.9 \pm 1.3$ \\
\tableline
\end{tabular}
\tablecomments{Frequencies in $\mu$Hz of individual modes identified
in HD 17156.}
\end{center}
\end{table}

\begin{table}
\begin{center}
\caption{Stellar evolution models fitting the observed 
frequencies in Table~1.\label{tbl-2}}
\begin{tabular}{ccccccccccc}
\tableline\tableline
$No$ & $M_*/{\rm M}_\odot$ & Age & $Z_0$ & $X_0$ & $R_*/{\rm R}_\odot$ & 
$\rhomean$ & $T_{\rm eff}$ & $L_*/{\rm L}_\odot$ & $\chi_\nu^2$ & $\chi^2$ \\
     &                     & (Gyr)&      &       &                     &
$({\rm g\,cm^{-3}})$ & (K) &  &  &   \\
\tableline
1$^{\rm a}$ & 1.28 & 2.936 & 0.0299 & 0.6803 & 1.505 & 0.5292 & 6058 & 2.74 & 2.46 & 2.57 \\ 
2 & 1.29 & 2.756 & 0.0299 & 0.6803 & 1.508 & 0.5301 & 6079 & 2.79 & 2.43 & 2.43 \\ 
3$^{\rm a}$ & 1.30 & 2.557 & 0.0299 & 0.6803 & 1.509 & 0.5325 & 6123 & 2.87 & 2.59 & 3.14 \\ 
4 & 1.31 & 2.389 & 0.0299 & 0.6803 & 1.512 & 0.5337 & 6144 & 2.93 & 2.65 & 3.85 \\ 
5 & 1.32 & 2.215 & 0.0299 & 0.6803 & 1.514 & 0.5357 & 6176 & 2.99 & 2.82 & 5.47 \\ 
6 & 1.28 & 3.069 & 0.0338 & 0.6687 & 1.507 & 0.5264 & 5987 & 2.62 & 2.48 & 4.80 \\ 
7 & 1.29 & 2.865 & 0.0338 & 0.6687 & 1.510 & 0.5279 & 6021 & 2.69 & 2.40 & 3.31 \\ 
8 & 1.30 & 2.670 & 0.0338 & 0.6687 & 1.512 & 0.5294 & 6054 & 2.76 & 2.34 & 2.51 \\ 
9$^{\rm a}$ & 1.31 & 2.474 & 0.0338 & 0.6687 & 1.514 & 0.5316 & 6099 & 2.85 & 2.52 & 2.64 \\ 
10 & 1.32 & 2.307 & 0.0338 & 0.6687 & 1.517 & 0.5329 & 6120 & 2.90 & 2.56 & 3.03 \\ 
\tableline
\end{tabular}
\tablecomments{Models minimizing $\chi_\nu^2$ (cf.\ Eq.~\ref{eq:chisqnu})
along the evolution tracks illustrated in Fig.~\ref{fig15}.
The smallest value of $\chi_\nu^2$ is obtained for model 8.
Models are shown for two values of the initial heavy-element and
hydrogen abundances $Z_0$ and $X_0$, corresponding to
${\rm [Fe/H] = 0.24}$ and 0.30.
Models indicated by superscript `a' were computed with overshoot with
$\alpha_{\rm ov} = 0.05$, the remaining models had no overshoot.}
\end{center}
\end{table}

\begin{table}
\begin{center}
\caption{System Parameters of HD 17156.\label{tbl-3}}
\begin{tabular}{lccc}
\tableline\tableline
Parameter & Value & 1-$\sigma$ limits & Comment \\
\tableline
Mean density, $\rhomean$ (g\,cm$^{-3}$) & 0.5301 & $\pm$0.0044 & A \\
Age (Gyr) & 2.8 & $\pm$0.6 & A \\
Effective Temp, $T_{\rm eff}$(K) & 6082 & $\pm$60 & B \\
Surface gravity, $\log(g)$ (cgs) & 4.31 & $\pm$0.04 & B \\
Metallicity, [Fe/H]$^*$ & 0.24 & $\pm$0.03 & B \\
Mass, $M_*$ (${\rm M}_{\odot}$) & 1.285 & $\pm$0.026 & C \\
Radius, $R_*$ (${\rm R}_{\odot}$) & 1.507 & $\pm$0.012 & C \\
Age (Gyr) & 3.2 & $\pm$0.3 & C \\
Luminosity, $L_*$ (${\rm L}_{\odot}$) & 2.79 & $\pm$0.14 & D \\
Surface gravity, $\log(g)$ (cgs) & 4.191 & $\pm$0.004 & D \\
\tableline
\end{tabular}
\tablecomments{A:  Based on asteroseismic analysis of this paper.
B: Spectroscopic results averaging from \citet{fis07} and \citet{amm09}.
$^*$ For the purposes of deriving MCMC based errors \citep{bro09} on
$M_*$ and $R_*$ we have adopted a doubling of the [Fe/H] error to $\pm$ 0.06.
C: Function of A and B parameters and use of stellar evolution models.
D: Derived from B and C parameters.
}
\end{center}
\end{table}


\end{document}